\documentstyle[12pt] {article}
\begin {document}
\parindent=15pt
\begin{center}
\vskip 1.5 truecm
{\bf SOLUTION OF THE HIGH TWIST EVOLUTION EQUATION
IN THE DOUBLE LOGARITHMIC APPROXIMATION
IN QCD }\\
\vspace{1cm}
A.Shuvaev\\
Theory Department, St.Petersburg Nuclear Physics Institute\\
188350, Gatchina, St.Petersburg, Russia.
\end{center}
\begin{abstract}
The evolution equation for high twist operators is solved in the Double
Logarithmic Approximation for cylinder-type diagrams which dominate in the
limit of large number of colors.
The asymptotic behaviour of the parton correlation
function is shown to be determined by the spectrum of the Ising model for a
one-dimensional chain of spin-1/2 magnets.
\end{abstract}
\vspace{1cm}
\newpage
{\bf 1.} The strong growth of the deep-inelastic structure function with
decreasing Bjorken variable $x_B$ leads to the necessity for taking into
account the contribution of high twist operators since when $x_B\rightarrow
0 $ the power suppression $1/Q^2$ can be compensated by the large
logarithmic factors $\sim \ln 1/x_B$. Indeed, the moments of the structure
function
\begin{equation}
\label{1}M(j,Q^2)=\int_0^1dx_Bx_B^{j-1}x_BG(x_B,Q^2),
\end{equation}
where $Q^2$ is a square of virtual photon momentum, can be expressed through
a Wilson Operator Product Expansion in the form (see\cite{collins})
\begin{equation}
\label{2}M(j,Q^2)=\sum_{N\geq 0}\frac 1{Q^{2N}}C_{2N}(j,Q^2)\langle h\mid
O^{(2N)}\mid h\rangle .
\end{equation}
Here $C_{2N}(j,Q^2)$ is the coefficient function and $\langle h\mid
O^{(2N)}\mid h\rangle $ denotes the expectation value of the local operator
in a hadron state. Usually one can neglect the contribution of high twists
(i.e. the terms with $N\geq 1$ in (\ref{2})) as they are suppressed by the
powers of the large momentum square $Q^2$. However this is not the case for $
x_B\rightarrow 0$. As is known from renormalization group arguments the
coefficient function is expressed through anomalous dimension $\gamma
_{2N}(j-1)$
\begin{equation}
\label{3}C_{2N}(j,Q^2)\sim e^{\gamma _{2N}(j-1)\ln Q^2/\mu ^2},
\end{equation}
where $\mu $$^2$ is an infrared cut-off. The small-$x_B$ behavior of the
structure function is controlled by the singularity of $\gamma _{2N}$ at $
j=1 $ . For the twist 2 case the anomalous dimension is given by the kernel
of Gribov-Lipatov-Altarelli-Parisi (GLAP) equation in the gluon-gluon channel,
and near $j=1$ it has the form:
\begin{equation}
\label{3a}\gamma _2(j-1,Q^2)=\frac{N_c\alpha _S}{\pi (j-1)}
\end{equation}
($N_c$ is the number of colors). The quark contributions are negligible here
since they are not singular for $j\rightarrow 1$. Supposing the high twist
anomalous dimension $\gamma _{2N}$ to have the same singularity at $j=1$ one
can estimate the contribution of the twist $2N$ operator to the structure
function for $x_B\rightarrow 0$ as
\begin{equation}
\label{4}x_BG_{2N}(x_B,Q^2)\sim \left( \frac {\mu ^2}{Q^2}\right)^N\exp
\lambda _{2N} \sqrt{\ln Q^2/\mu ^2\ln 1/x_B},
\end{equation}
where the coefficient $\lambda $$_{2N}$ is determined by the residue of $
\gamma _{2N}$ at $j=1$ .This relation suggests existence of a kinematic
region of the $x_B,Q^2$ variables where the power suppression is compensated
by the logarithmic factors in the exponent. The line where the twist 2 and
twist 4 contributions are of the same order is the boundary of this region,
and in the interior of it the twist 4 becomes leading.
There is an infinite set of similar
domains where the twist 6 operators turn out to be dominant ones, then twist
8 and so on. The exact location and boundaries of these domains are determined
by the coefficients $\lambda $$_{2N}$ in Eq.(\ref{4}).Inside these domains
GLAP evolution equation becomes invalid and has to be replaced by the
appropriate equation for high twists.

In paper \cite{glr} the specific contribution to the anomalous dimension $
\gamma _{2N}$ coming from the exchange of $2N$ non-interacting ladders
(Pomerons) was found to be
\begin{equation}
\label{5}
\begin{array}{cc}
\gamma _{2N}(\omega )=N\gamma _2(\frac \omega N), & \omega =j-1
\end{array}
\end{equation}
or
\begin{equation}\label{Pom}
x_BG_{2N}(x_B,Q^2)\sim \left( \frac {\mu ^2}{Q^2}\right)^N\exp N
\sqrt{\frac{\alpha_S N_c}{4\pi}\ln Q^2/\mu ^2\ln 1/x_B}
\end{equation}
that is the product of one-Pomeron asymptotics.

The calculations for twist 4 case performed in Refs.\cite{bartels,lrsh}
with the account of Pomerons interaction in Double Logarithmic
Approximation (DLA) modified this result:
\begin{equation}
\label{6}\gamma _4(\omega )=2\gamma _2(\frac \omega 2)(1+\delta )=\frac{
4N_c\alpha _S}{\pi \omega }(1+\delta )
\end{equation}
with $\delta \simeq (N_c^2-1)^2\approx 10^{-2}.$ The corrections due to
other states (color ladders) were estimated to be approximately of the
same order. The $\delta $
value which indicates the deviation from Eq.(\ref{5}) can be treated as a
measure of the Reggeon interaction strenght.
The problem how the Reggeon interaction modifies result (\ref{Pom}) will
be concidered below.

{\bf 2.} This paper will be based on the direct
solution of the evolution equation for high twist operators in DLA.

The evolution equation for the high twist operators was derived in paper
\cite{lipatov}. It generalizes the GLAP equation for quasipartonic operators
which form the closed set of high twist operators allowing for the
interpretation in terms of the parton model. Decomposing the momenta of
virtual gluons $k_i$ into Sudakov variables
$$
\begin{array}{ccc}
k_i=\alpha _iq^{\prime }+\beta _ip^{\prime }+k_{\bot i}, & (p^{\prime
})^2=(q^{\prime })^2=0, & k_{\bot i}p^{\prime }=k_{\bot i}q^{\prime }=0,
\end{array}
$$
where $p^{\prime }$ is a hadron momentum, $q=q^{\prime }-x_Bp^{\prime }$ is
a photon momentum, $q^2=-Q^2$, matrix elements of the twist $N$
quasipartonic operators between parton states is expressed only through the $
\beta _i$ variables and has a general form
\begin{equation}\label{7a}
\rho_{0,\mu _1,\ldots ,\mu _{N}}^{a_1,\ldots ,a_{N}}(\beta_1,\ldots,\beta_N)
\,=\,\Gamma _{\mu _1,\ldots ,\mu _{N}}^{a_1,\ldots ,a_{N}}\beta
_1^{n_1}\cdots \beta _{N}^{n_{N}},
\end{equation}
where $n_i$ are integer numbers and
tensor $\Gamma $ specifies the color and Lorentz structure of the operator.
The variable $\beta _i$ has a meaning of the fraction of the total hadron
momentum carried by a parton.

In axial gauge $q_\mu ^{\prime }A_\mu =0$ only the ladder-type diagrams
contribute to the leading logarithms. For the twist $N$ case they comprise
the local operator vertex of the form (\ref{7a}) and $N$ gluons in $t$
-channel interacting through all possible $s$-channel gluons rungs. Very
important property of Leading Logarithmic Approximation (LLA) is strict
ordering of transverse momenta along the ladder graph, namely, the
transverse virtualities decrease from the top of the diagram to the bottom.
The range of integration in each ladder cell is such that transverse
momentum above the rung is greater than the momentum below it. In other
words the above momentum plays the role of ultraviolet cut-off for the below
cell. Thus the $Q^2$ value, being the greatest momentum in the upper loop
that is attached to the local operator, is the ultraviolet cut-off for the
whole graph. The evolution equation for the quasipartonic operators is
derived from the diagrams with pure longitudinal external gluons' momenta $
p_i\simeq \beta _ip^{\prime }$. Their sum determines the parton correlation
function $\rho (Q^2,\beta _1,\cdots ,\beta _{N})$ the evolution equation
for which is obtained by taking the derivative of the diagrams with respect
to $\ln Q^2$. It has the form of the $N$-particle one dimensional
equation with pairwise interaction in which the variable $
\xi (Q^2)\equiv \int_{\mu ^2}^{Q^2}\frac{dk^2}{k^2}\frac{\alpha _S(k^2)}{
4\pi }$ is an effective ''time'' ($\alpha _S(k^2)$ is the running coupling
constant)
\begin{equation}
\label{8}\frac \partial {\partial \xi }\rho (\xi ,\beta _1,\cdots ,\beta
_N)\,=\,\sum_{i<j}\int d\beta _i^{\prime }d\beta _j^{\prime }\,\delta (\beta
_i+\beta _j-\beta _i^{\prime }-\beta _j^{\prime })\times
\end{equation}
$$
\times \Phi (\beta _i,\beta _j\mid \beta _i^{\prime },\beta _j^{\prime
})\,\rho (\xi ,\beta _1,\cdots ,\beta _i^{\prime },\cdots ,\beta _j^{\prime
},\cdots ,\beta _N)
$$
(to be more compact the color and Lorentz indexes are suppressed here). The
kernel $\Phi (\beta _i,\beta _j\mid \beta _i^{\prime },\beta _j^{\prime })$
is determined by the logarithmic part of the one-loop integral over $k_{\bot
}$ and $\alpha $. The initial condition for the function $\rho (\xi )$ is
given by the expression (\ref{7a}) that corresponds to the absence of loops.
The equation (\ref{8}) collects all the powers of $\alpha _S\ln Q^2/\mu ^2$
where the infrared cut-off $\mu ^2\ll Q^2$ is the characteristic hadron
scale.

The longitudinal momenta $\beta _i$ are not in general ordered in LLA. But
in DLA they also have to be ordered to provide a large logarithm for each
ladder cell. The $\beta _i$ variables increase from the values $\sim x_B$ at
the local operator vertex to the order of unity ones in the lower part of a
diagram. In such a kinematic the logarithmic divergencies that occur in
every loop when $\beta _i^{\prime }\rightarrow 0$ is cut from below by a
longitudinal momentum in the upper cell. Thus DLA implies the loop
integration in the evolution equation (\ref{8}) to be limited by the
condition
\begin{equation}
\label{9}\beta _i^{\prime },\beta _j^{\prime }\ll \beta _i,\beta _j
\end{equation}
which means that the momenta below the $s$-channel rung ($\beta $) and above
it ($\beta ^{\prime }$) are of different order of magnitude. The most
singular contribution comes in Eq.(\ref{8}) from the region where both $
\beta _i^{\prime }$ and $\beta _j^{\prime }$ tend to zero. The momentum
conservation allows it only if
\begin{equation}
\label{10}\beta _i+\beta _j\ll \beta _i,\beta _j
\end{equation}
that is with logarithmic accuracy $\beta _i\approx -\beta _j$. Hereafter it
is convenient to assume the momenta directed upward to be positive, directed
downward to be negative.

In the logarithmic domain the kernel of the evolution equation can be easily
obtained in helicity basis \cite{lipatov} by keeping the most singular in $
\beta ^{\prime }$ terms in gluon-gluon channel and integrating it over $
k_{\bot }$ and $\alpha $
\begin{equation}
\label{11}\Phi _{\mu ,\nu ,\lambda ,\sigma }^{a,b,c,d}(\beta _1,\beta _2\mid
\beta _1^{\prime },\beta _2^{\prime })\,=\,2\delta _{\mu ,\nu }\delta
_{\lambda ,\sigma }if^{a,c,g}if^{b,g,d}\,\beta _1\delta (\beta _1-\beta
_2)\,\frac 1{\beta _1^{\prime }\beta _2^{\prime }}.
\end{equation}
Here $\mu ,\nu ,\lambda ,\sigma $ are the two dimensional transverse indexes
, $f^{a,c,g}$ are the structure constants of the $SU(N_c)$ group. The
momenta $\beta _1,\beta _2$, are equal in DLA . They are positive but have
the opposite directions, one of them is incoming from below the other is
outgoing. The low-scale momenta $\beta _1^{\prime }$ and $\beta _2^{\prime }$
are not supposed to be equal in DLA since the momentum transfer from below $
\beta _1-\beta _2$ is small only compared to the large momenta $\beta _1$
and $\beta _2$ but is of the same order as the low-scale ones. However $
\beta _1^{\prime }$ and $\beta _2^{\prime }$ like $\beta _1$ and $\beta _2$
have the opposite directions, so the incoming from below momentum turns out
into outgoing and vice versa. This property of the longitudinal momenta
directions can be formulated as an ''arrow rule''\cite{manukian}. One can
assign an arrow to each $t$-channel gluon line to indicate the momentum
direction. The interaction occurs only between the lines with the opposite
arrows while for the particular line the arrow direction does not change
through all interactions. As a consequence all $t$-channel lines are divided
into those incoming the local operator vertex and outgoing it. The violation
of the ''arrow rule'' results in the loss of the longitudinal logarithm in
the cell integration producing only small factor $\sim \alpha _S$.
One should
note here that the ''arrow rule'' will be effectively absorbed into
restrictions imposed in the $N_c\rightarrow \infty $ limit where only the
''neighbouring'' gluons are allowed to interact.

It should be noted that the restriction only to the ladder type diagrams is
valid in DLA provided the colorless in $t$-channel states are considered.
Indeed, for these states only hard $s$-channel gluons carrying the
longitudinal momentum fraction $\beta ^{(s)}=\beta _i-\beta _i^{\prime }\sim
\beta _i$ are emitted. The logarithmic contributions from the soft
non-ladder gluons emission cancel out when all the diagrams are summed up,
since the soft particle probes the total color charge of the system which is
zero for white operators \cite{lrsh}.

The kernel (\ref{11}) gives rise to DLA evolution equation
$$
\frac \partial {\partial \xi }\rho _{\mu _1,\ldots ,\mu _{N};\sigma
_1,\ldots ,\sigma _{N}}^{a_1,\ldots ,a_{N}}(\beta _1,\ldots ,\beta _{N})=
$$
\begin{equation}
\label{12}=\sum_{k<j}2S_{\sigma _k,\sigma _j}\delta _{\mu _k,\mu _j}\delta
_{\lambda _k,\lambda _j}if^{a_k,b_k,g}if^{g,b_j,a_j}\,\beta _k\delta (\beta
_k-\beta _j)\times
\end{equation}
$$
\times \int_0^{\beta _k}\frac{d\beta _k^{\prime }}{\beta _k^{\prime }}
\,\int_0^{\beta _j}\frac{d\beta _j^{\prime }}{\beta _j^{\prime }}\,\rho
_{\mu _1,\ldots ,\mu _{N};\sigma _1,\ldots ,\sigma _{N}}^{a_1,\ldots
,b_k,\ldots ,b_j,\ldots ,a_{N}}(\beta _1,\ldots ,\beta _k^{\prime },\ldots
,\beta _j^{\prime },\ldots ,\beta _{N}).
$$
Here all $\beta _i>0$, the momentum direction and the ''arrow rule''
are specified by the indexes
$\sigma $ and coefficient $S_{\sigma _k,\sigma _j}$. The integration
range in Eq.(\ref{12}) is determined with logarithmic accuracy by condition
( \ref{9}). The initial condition is given by Eq.(\ref{7a})

For twist-2 colorless operators the solution to Eq.(\ref{12}) has the form
$$
\rho _{\mu _1,\mu _2}^{a_1,a_2}(\xi ,\beta _1,\beta _2)=\delta
_{\mu _1,\mu _2}\delta ^{a_1,a_2}\delta (\beta _1-\beta _2)
f(\xi ,\frac{x_B}\beta ),
$$
where function $f$ obeys conventional GLAP evolution equation
after substitution $\beta =1$ (the structure function is considered to
be dependent on the variable $-\frac{q^2}{2pq}=\frac{x_B}\beta $).

The color structure of the evolution equation is essentially simplified
in the $N_c\rightarrow \infty $ limit.
If in this case the color index is written as the pair of quark and
antiquark ones $a=(i,\overline{\j})$ and gluons are drawn by double lines the
interaction occurs only between the
gluons containing the common ''quark'' line. The closed cycle arising when
such $t$-channel gluon lines are connected produces a factor $\frac 12N_c$
while joining other lines gives an order of unity factor. There is a
significant difference here with the twist-2 (Pomeron) solution for which
the factor $N_c$ is obtained after adding together two possible joining of
two $t$-channel gluon lines.
The gluons can be enumerated in such a way that the gluon with number $n$
is allowed to interact only with the gluons $n-1$ and $n+1$. All the gluons
can be represented to be lying on the cylinder in the color space and each
of them interacts only with its neigbours. It is natural to assign to the
neighbouring gluons opposite signs of $\beta$ to ensure interaction between
them. After this the color structure of the function $\rho$
turns into itself under the action of the kernel of the evolution equation
which takes the form
$$
\frac \partial {\partial \xi }\rho _{\mu _1,\ldots ,\mu _{N}}
(\beta _1,\ldots ,\beta _{N})=
$$
\begin{equation}\label{ee}
=\sum_{n}\delta _{\mu_{n},\mu_{n+1}}\delta
_{\mu_{n}^{\prime},\mu_{n+1}^{\prime}}\,\beta _n\delta (\beta
_n-\beta _{n+1})\times
\end{equation}
$$
\times \int_0^{\beta _n}\frac{d\beta _n^{\prime }}{\beta _n^{\prime }}
\,\int_0^{\beta _{n+1}}\frac{d\beta _{n+1}^{\prime }}{\beta _{n+1}^{\prime }}
\,\rho
_{\mu _1,\ldots ,\mu_n^{\prime },\mu_{n+1}^{\prime },\ldots, \mu _{N}}
(\beta _1,\ldots ,\beta _n^{\prime },
\beta _{n+1}^{\prime },\ldots ,\beta _{N}).
$$
Here $\xi$ is redefined (for large $N_c$ case) as
\begin{equation}\label{xi}
\xi (Q^2)\,=\, \int_{\mu ^2}^{Q^2}\frac{dk^2}{k^2}\frac{N_c \alpha _S(k^2)}{
4\pi }.
\end{equation}

{\bf 3.} There is a way to transform evolution equations (\ref{12}) or
(\ref{ee}) to the Schr\"odinger-type differential equation.
The Laplace transformation
\begin{equation}
\label{18}\rho (E)\,=\,\int_0^\infty d\xi \,e^{-E\xi }\rho (\xi )
\end{equation}
brings the evolution equation to the stationary form, which in the
logarithmic coordinates $x_i=\ln 1/\beta _i$ reads
\begin{equation}
\label{19}E\rho _{r_1,\ldots ,r_{N}}(x_1,\ldots ,x_{N})=
\end{equation}
$$
=-\sum_{i<k}\delta (x_i-x_k)\,\int_{x_i}^\infty dx_i^{\prime
}\,\int_{x_k}^\infty dx_k^{\prime }\,A_{ik}\,\rho _{r_1,\ldots
,r_{N}}(x_1,\ldots ,x_i^{\prime },\ldots ,x_k^{\prime },\ldots ,x_{N}).
$$
The arrow,Lorentz and color indexes are accumulated here in the index $
r=\{\sigma ,\mu ,a\}$ and the matrix $A_{ik}$ acting on the indexes at $i,k$
locations is introduced
\begin{equation}
\label{20}A_{ik}=2S_{\sigma _i,\sigma _k}\delta _{\mu _i,\mu _k}\delta _{\mu
_i^{\prime },\mu _k^{\prime }}if^{a_i,a_i^{\prime },g}if^{g,a_k^{\prime
},a_k}.
\end{equation}

Seeking the solution of Eq.(\ref{19}) in the form
\begin{equation}
\label{21}\rho _{r_1,\ldots ,r_{N}}(x_1,\ldots ,x_{N})=\frac \partial
{\partial x_1}\ldots \frac \partial {\partial x_{N}}\,\varphi _{r_1,\ldots
,r_{N}}(x_1,\ldots ,x_{N})
\end{equation}
the function $\varphi $ obeys the equation
\begin{equation}
\label{22}E\prod_{i=1}^{N}\frac \partial {\partial x_i}\,\varphi
_{r_1,\ldots ,r_{N}}(x_1,\ldots ,x_{N})=
\end{equation}
$$
=-\sum_{i<k}\delta (x_i-x_k)\prod_{
\begin{array}{c}
j\neq i \\
j\neq k
\end{array}
}\frac \partial {\partial x_j}\,A_{ik}\,\varphi _{r_1,\ldots
,r_{N}}(x_1,\ldots ,x_{N})
$$
($\varphi $ is assumed to be vanishing at the upper limit). The coordinate
space is separated into $N!$ sectors which are obtained from the
fundamental one
\begin{equation}
\label{23}x_1<x_2<\cdots <x_{N-1}<x_{N}
\end{equation}
by the permutation of any two variables. The function $\varphi $ satisfies
inside these sectors the free equation
\begin{equation}
\label{24}E\prod_{i=1}^{N}\frac \partial {\partial x_i}\,\varphi
_{r_1,\ldots ,r_{N}}(x_1,\ldots ,x_{N})=0
\end{equation}
and has the form
\begin{equation}
\label{25}\varphi _{r_1,\ldots ,r_{N}}(x_1,\ldots
,x_{N})\,=\,e^{k_1x_1+\cdots +k_{N}x_{N}}
\end{equation}
or linear combination of such terms with some $k_i$ equal to zero in
each of them. At the boundary separating the two sectors the matching
condition has to be fulfilled. Near, say, $x_1=x_2$ hyper plane it reads
\begin{equation}
\label{26}\prod_{j=3}^{N}\frac \partial {\partial x_j}E\,\frac \partial
{\partial x_1}\frac \partial {\partial x_2}\,\varphi _{r_1,\ldots
,r_{N}}=-\prod_{j=3}^{N}\frac \partial {\partial x_j}\,\delta
(x_1-x_2)\,\varphi _{r_1,\ldots ,r_{N}}.
\end{equation}
Denoting the solution in the sectors $x_1>x_2$ and $x_1<x_2$ as $\psi ^{12}$
and $\psi ^{21}$ respectively, so that
\begin{equation}
\label{27}\varphi =\theta (x_1-x_2)\psi _{_{r_1,\ldots
,r_{N}}}^{12}(x_1,\ldots ,x_{N})+\theta (x_2-x_1)\psi _{r_1,\ldots
,r_{N}}^{21}(x_1,\ldots ,x_{N})
\end{equation}
the functions $\psi $ are connected through the relations
\begin{equation}
\label{28}
\begin{array}{cc}
1) & \psi ^{12}\mid _{x_1=x_2}=\psi ^{21}\mid _{x_1=x_2} \\
2) & E\left. \left[ \left( \frac \partial {\partial x_1}\psi ^{21}-\frac
\partial {\partial x_1}\psi ^{12}\right) +\left( \frac \partial {\partial
x_2}\psi ^{12}-\frac \partial {\partial x_2}\psi ^{21}\right) \right]
\right| _{x_1=x_2}= \\
& =\left. -A_{12}\left( \psi ^{12}+\psi ^{21}\right) \right| _{x_1=x_2}.
\end{array}
\end{equation}

Conditions (\ref{28}) allow to establish the connection between the
evolution equation and one-dimensional quantum mechanics of $N$ particles
described by the Schr\"odinger equation
\begin{equation}
\label{29}\frac E2\sum_j\frac{\partial ^2}{\partial x_j^2}\,\varphi
_{r_1,\ldots ,r_{N}}-\sum_{i<k}\delta (x_i-x_k)\,A_{ik}\,\varphi
_{r_1,\ldots ,r_{N}}\,=\,{\cal E}\varphi _{r_1,\ldots ,r_{N}}.
\end{equation}
Indeed the function $\varphi $ clearly has form (\ref{25}) inside any sector
of type (\ref{23}) and satisfies conditions (\ref{28}) at its boundary. Thus
Eqs.(\ref{22}) and (\ref{29}) have the same wavefunction although there is
not straightforward connection between the energies $E$ and ${\cal E.}$
For the large $N_c$ case Eq.(\ref{29}) can be rewritten in a chain-like form
\begin{equation}\label{cc}
\frac E2 \sum_n {\cal H}_{n,n+1}\varphi _{\mu_1,\ldots ,\mu_{N}}
(x_1,\ldots ,x_{N})\,=\,{\cal E}\varphi _{\mu_1,\ldots ,\mu_{N}}
(x_1,\ldots ,x_{N}),
\end{equation}
$$
{\cal H}_{i,k}\,=\,\frac 12 \bigl(\frac{\partial^2}{\partial x_i^2}\,+\,
\frac{\partial^2}{\partial x_k^2} \bigr)\,+\,\delta (x_i - x_k)
\delta_{\mu_i \mu_k}
\delta_{\mu_i^{\prime} \mu_k^{\prime}}.
$$

The correlation function $\rho $ built upon function $\varphi$
through (\ref{21}) represents the sum of the terms with various
number of $\delta $-functions. Note that
the term without $\delta $-functions vanishes after some $k_i$ are put to
zero. Its absence is evident also from the structure of the equation (\ref
{22}).

{\bf 4.} Instead of the solution of evolution equation (\ref{ee})
or (\ref{cc}) the approach based on the direct summation of
the ladder-type diagrams in the $N_c \rightarrow \infty$ limit will
be adopted here. To begin with the formal solution to the evolution equation
(\ref{8}) can be written as
\begin{equation}
\label{fsol}\rho (\xi )=e^{\Phi \xi }\rho _0=\sum_{n=0}^\infty \frac
1{n!}\,\xi ^n\Phi ^n\rho _0,
\end{equation}
where $\Phi $ is the kernel of the evolution equation. The term $\Phi ^n$
acting on the initial state $\rho _0$ generates the ladder-type diagrams
with $n$ loops in $\beta $-space (with $n$ integrals over $\beta $). To
restore the proper expression (\ref{fsol}) each $\beta $-loop has to be
multiplied by variable $\xi $ and the whole diagram has to be divided by $n!$
. Introducing function
$$
W(E)\,=\,\sum_{n=0}^\infty \frac 1{E^n}\,\Phi ^n\rho _0
$$
the solution (\ref{fsol}) is given by the formula
$$
\rho (\xi )=\frac 1{2\pi i}\,\oint \frac{dE}E\,e^{E\xi }W(E),
$$
where the integration goes aroung $E=0$ point in the counter-clock-wise
direction. Function $W(E)$ obeys Bethe-Solpiter equation
\begin{equation}
\label{BS}W=\rho _0\,+\,\frac 1E\,\Phi W
\end{equation}
describing the system of $N$ $t$-channel particles with pairwise
interaction between particles $n$ and $n+1$
\begin{equation}
\label{kern1}\Phi_{\mu ,\nu ,\lambda ,\sigma }(\beta _1,\beta _2\mid
\beta _1^{\prime },\beta _2^{\prime })\,=\,\delta _{\mu ,\nu
}\delta _{\lambda ,\sigma }\,\beta _1\delta (\beta _1-\beta _2)\,\frac
1{\beta _1^{\prime }\beta _2^{\prime }}.
\end{equation}
Equation (\ref{BS}) gives rise to the set of ladder-type planar diagrams in $
\beta $-space. To sum it up they can be transformed to the form where the
interaction between $t$-channel gluons is treated as an exchange of $s$
-channel particles. Kernel (\ref{kern1}) results effectively from the
exchange of the particle whose emission is given by the expression (Fig.1):
\begin{equation}
\label{ampl}T_{\mu \nu }\,=\,\frac 1{\sqrt{E}}\,
\theta(\beta-\beta^{\prime})\frac 1{\beta
^{\prime }}\,\sqrt{\beta }\,a_{\mu \nu }^{\dagger }(\beta ),
\end{equation}
where $a_{\mu \nu }^{\dagger }(\beta )$ ($a_{\lambda \sigma }(\beta )$) is a
creation (annihilation) operator
$$
\left[ a_{\lambda \sigma }(\beta )\,,\,a_{\mu \nu }^{\dagger }(\beta
^{\prime })\right] =\delta _{\lambda \mu }\delta _{\sigma \nu }\delta (\beta
-\beta ^{\prime }).
$$
Indeed, amplitude (\ref{ampl}), being squared, reproduces evidently (\ref
{kern1}).

In a ladder-type diagram any $t$-channel line emits and absorbs arbitrary
number of $s$-channel particles. For the particular $t$-channel line with
number $n$ there is associated an amplitude containing creation operators
$a_{n,\mu \nu }^{\dagger }(\beta )$ and annihilation ones $a_{n-1,\lambda
\sigma }(\beta )$. For instance, the amplitude shown in Fig.2 has a form
$$
T_{\mu _1\mu _6}\,=\,E^{-\frac 52}\,
\sum_{\mu _2,\cdots ,\mu _5}\int dx\beta _1\cdots d\beta
_4\times
$$
$$
\times \sqrt{x}\frac{a_{n,\mu _1\mu _2}(x)}{\beta _1
}\,\sqrt{\beta _1}\frac{a^{\dagger }_{n-1,\mu _2\mu _3}(\beta _1)}{\beta _2}
\,\sqrt{
\beta _2}\frac{a_{n,\mu _3\mu _4}(\beta _2)}{\beta _3}\times
$$
$$
\times \sqrt{\beta _3}\frac{a_{n,\mu _4\mu _5}(\beta _3)}{\beta _4
}\,\sqrt{\beta _4}\frac{a^{\dagger }_{n-1,\mu _5\mu _6}(\beta _4)}{y},
$$
where the integration over the longitudinal momenta is submitted to the
restriction
\begin{equation}
\label{order}x>\beta _1>\beta _2>\beta _3>\beta _4>y.
\end{equation}
Multiplying expressions of this type and applying Wick theorem to the
product of the creation and annihilation operators one can reproduce
all the ladder
diagrams for the equation (\ref{BS}). Note that
only the planar graphs will appear since the diagrams with the crosses
between $s$-channel particles destroy $\beta $ ordering (\ref{order}).
Assembling for a given $t$-channel line the terms with all possible
emissions and absorbtions one gets the expression
$$
T_{\mu \mu ^{\prime }}^{(n)}(x,y)\,=\,
\sum_k E^{-\frac k2}\,\sum_{\mu _1,\cdots ,\mu
_k}\int_{y}^xd\beta _1\int_{y}^{\beta _1}d\beta _2\cdots
\int_{y}^{\beta _{k-1}}d\beta _k\times
$$
\begin{equation}
\label{Tk}\times \frac{\sqrt{x}}{\beta _1}\,\left( a_{n,\mu \mu _1}
(x)+a^{\dagger }_{n-1,\mu \mu _1}(x)\right) \,
\frac{\sqrt{\beta _1}}{\beta _2}\,\left(
a_{n,\mu _1\mu _2}(\beta _1)+a^{\dagger }_{n-1,\mu _1\mu _2}(\beta
_1)\right)  \times
\end{equation}
$$
\cdots\times \frac{\sqrt{\beta _k}}{y}\,\left( a_{n,\mu _k\mu ^{\prime
}}(\beta _k)+a^{\dagger }_{n-1,\mu _k\mu ^{\prime }}(\beta _k)\right) .
$$
Here $x$ and $y$ are the momenta of lower and upper points of the
$t$-channel line (incoming and outgoing momenta, see Fig.2), $\mu $ and $\mu
^{\prime }$ are their polarization indexes. Introducing $2\times 2$ matrix $
A_n(\beta )$
$$
A_{n,\mu \nu }(\beta )\,=\,a_{_{n,\mu \nu }}(\beta )+a^{\dagger }_{n-1,\mu
\nu }(\beta )
$$
Eq.(\ref{Tk}) can be rewritten as
$$
T_{\mu \mu ^{\prime }}^{(n)}(x,y)=
$$
$$
=\sum_k\left(\frac 1{\sqrt{E}}\, \frac{\sqrt{x}}{y}A_n(x)\,
\int_{y}^xd\beta
_1\int_{y}^{\beta _1}d\beta _2\cdots \int_{y}^{\beta _{k-1}}d\beta _k\,
\frac{A_n(\beta _1)}{\sqrt{E\beta _1}}\,
\cdots \frac{A_n(\beta _k)}{\sqrt{E\beta _k}}\right) _{\mu \mu ^{\prime }}=
$$
\begin{equation}
\label{Tn}
=\left(  \frac{\sqrt{x}}{x_B}\,A_n(x)\,P\exp \left\{ \int_{y}^xd\beta
\,\frac{
A_n(\beta )}{\sqrt{E\beta}} \right\} \right) _{\mu \mu ^{\prime }}.
\end{equation}
Here symbol $P$ means ordering of $A_n(\beta )$ matrixes according which the
matrix with larger argument stands to the left from the matrix with smaller
one. An equivalent form of the latter expression is
\begin{equation}\label{dP}
T_{\mu \mu ^{\prime }}^{(n)}(x,x_B)=
\sqrt{E}\frac{x}{y}\,\frac{\partial}{\partial x}
\left(P\exp \left\{ \int_{y}^xd\beta \,\frac{
A_n(\beta )}{\sqrt{E\beta}} \right\} \right) _{\mu \mu ^{\prime }}.
\end{equation}

Now the problem reduces formally to the evaluation of the correlator
$$
K_{\mu_1,\mu_1^{\prime},\ldots,\mu_N,\mu_N^{\prime}}(x_1,y_1,\ldots,x_N,y_N)=
$$
\begin{equation}\label{corr}
=\left\langle 0\left| T_{\mu _1\mu _1^{\prime }}^{(1)}(x_1,y_1)T_{\mu _2\mu
_2^{\prime }}^{(2)}(x_2,y_2)\cdots T_{\mu _N\mu _N^{\prime
}}^{(N)}(x_N,y_N)\right| 0\right\rangle ,
\end{equation}
where $\left| 0\right\rangle $ is the vacuum of operators $a_1,a_2,\ldots,
a_N $. Indeed, only Wick pairing terms survive after vacuum averaging,
therefore correlator (\ref{corr}) includes all the ladder diagrams for
Bethe-Solpiter equation (\ref{BS}), whose solution is expressed through
it as
$$
W_{\mu_1,\ldots,\mu_N}(x_1,\ldots,x_N)=
$$
$$
=\sum_{\mu^{\prime}}\int dy_1 \cdots dy_N\,
K_{\mu_1,\mu_1^{\prime},\ldots,\mu_N,\mu_N^{\prime}}(x_1,y_1,\ldots,x_N,y_N)
\rho_{0,\mu_1^{\prime},\ldots,\mu_N^{\prime}}(y_1,\ldots,y_N).
$$
For the cylinder topology case one has to replace in (\ref{corr})
$a^{\dagger}_0
\rightarrow a_N$ in the first term and $a_N\rightarrow a^{\dagger}_N$
in the last one. It is needed to ensure Wick pairing between $t$-channel
lines $1$ and $N$ to reproduce their interaction on the cylinder.

If it were not for the matrix structure of $T^{(n)}$ the correlator
would be easily calculated using the relation
\begin{equation} \label{uv}
\exp \{\int d\beta \,U(\beta )\,a_n(\beta )\}\,\exp \{\int d\beta \,V(\beta
)\,a_n^{\dagger }(\beta )\}=
\end{equation}
$$
=\exp \{\int d\beta \,V(\beta )\,a_n^{\dagger }(\beta )\}\,\exp
\{\int d\beta \,U(\beta )\,a_n(\beta )\}\,\exp \{\int d\beta \,U(\beta
)V(\beta )\}
$$
which is valid for arbitrary functions $U(\beta)$ and $V(\beta)$.
Indeed, in that case the correlator would be of the form
$$
K(x_1,y_1,\ldots,x_N,y_N)=
E^{\frac N2}\,\frac{x_1}{y_1}\cdots\frac{x_N}{y_N}\,
\frac{\partial}{\partial x_1}\cdots\frac{\partial}{\partial x_N}
\varphi(x_1,y_1,\ldots,x_N,y_N),
$$
where
$$
\varphi(x_1,y_1\ldots,x_N,y_N)=
$$
\begin{equation}\label{exps}
=\left\langle 0\left|\exp\left\{\int_{y_1}^{x_1} \frac{
d\beta}{\sqrt{E\beta}}a_1(\beta)\right\}\,\exp\left\{\int_{y_1}^{x_1} \frac{
d\beta}{\sqrt{E\beta}}a_n(\beta)\right\}\,\times \right. \right.
\end{equation}
$$
\left. \left. \times \exp\left\{\int_{y_2}^{x_2} \frac{
d\beta}{\sqrt{E\beta}}a_2(\beta)\right\}\,\exp\left\{\int_{y_2}^{x_2} \frac{
d\beta}{\sqrt{E\beta}}a^{\dagger}_1(\beta)\right\}\,\cdots\,
\right| 0\right\rangle.
$$
The evaluation of this expression can proceed as follows.
To begin with one can get
rid off the fourth exponent in (\ref{exps}) moving it to the left through
third and second ones, commuting it through
relation (\ref{uv}) with the first one and taking into account that
$\langle 0 \vert a^{\dagger}_1(\beta)=0$. Then the same can be done with the
other exponents moving the terms with creation operators to the left and with
annihilation ones to the right.
After a sequence of such manipulations the final result will be
\begin{equation}\label{fr}
\varphi(x_1,y_1,\ldots,x_N,y_N)=
\end{equation}
$$
=\exp\left\{\int_{y_{12}}^{x_{12}} \frac{
d\beta}{E\beta}+\int_{y_{23}}^{x_{23}} \frac{
d\beta}{E\beta}+\cdots+\int_{x_{N-1,N}}^{x_{N-1,N}} \frac{
d\beta}{E\beta} +\int_{x_{N,1}}^{x_{N,1}} \frac{
d\beta}{E\beta} \right\},
$$
where the integrations limits $x_{ik}=min\{x_i,x_k\}$ and
$y_{ik}=max\{y_i,y_k\}$ arise after
substitution in relation (\ref{uv}) $U_i(\beta)=V_i(\beta)=\frac 1
{\sqrt{E\beta}}\,\theta(x_i-\beta)\theta(\beta-y_i)$. Differentiating
these terms produces
pre-exponent factors $\delta(x_{i-1}-x_i)$. Their presence in the correlation
function has the origin in the evolution equation structure (see
Eqs.(\ref{21}) and (\ref{27})).
In DLA framework all $x_i$ in the exponents
have to be taken to be the same and all $y_i\simeq x_B$.
Then the leading behaviour of the correlator
will be
\begin{equation}\label{simp}
\left\langle 0\left| T^{(1)}(x,x_B)\cdots T^{(N)}(x,x_B)
\right| 0\right\rangle \sim \exp \{N\,\frac 1E \ln \frac
x{x_B}\}.
\end{equation}

{\bf 5.} The matrix structure of (\ref{Tn}) prevents
to apply relation (\ref{uv})
immediatly. To avoid this difficulty $P$-exponent in (\ref{Tn}) can be
represented through the functional integral. This expression is derived
in two steps. At first the $\beta $ interval is divided into large number $N$
of small intervals $\Delta \beta $:
\begin{equation}\label{prod}
P\,\exp \left\{ \int_{x_B}^x\frac{d\beta }{\sqrt{E\beta }}\,A(\beta )\right\}
\simeq
\end{equation}
$$\simeq \left( I+\frac{\Delta \beta }{\sqrt{E\beta _N}}\,A(\beta
_N)\right) \,\left( I+\frac{\Delta \beta }{\sqrt{E\beta _{N-1}}}\,A(\beta
_{N-1})\right) \cdots \left( I+\frac{\Delta \beta }{\sqrt{E\beta _0}}
\,A(\beta _0)\right)
$$
$$
\begin{array}{cc}
\beta _N=x & \beta _0=y,
\end{array}
$$
the accuracy being the better the smaller value $\Delta \beta $ is taken. At
the second step the following expression for the unit $2\times 2$ matrix is
inserted between the terms in the product (\ref{prod})
$$
c\int dz^{*}dz\,\delta (z^{*}z-1)\,z_\lambda z_\sigma ^{*}=
\delta _{\lambda \sigma }.
$$
Here $c$ is a normalization constant and
the integration is carried out over the components of two dimensional
complex vector $z$,
$$
\begin{array}{c}
dz^{*}dz\,=\,dz_2^{*}dz_2\,dz_1^{*}dz_1\\
\delta (z^{*}z-1)\,=\,\delta (z_2^{*}z_2+z_1^{*}z_1-1).
\end{array}
$$
It gives
$$
P\,\exp \left\{ \int_{x_B}^x\frac{d\beta }{\sqrt{E\beta }}\,A(\beta )\right\}
_{\mu \mu ^{\prime }}\simeq
\int \prod_{k=0}^{N+1}dz_k^{*}dz_k\delta (z_k^{*}z_k-1)\,z_{N+1,\mu}z^{*}_
{0,\mu^{\prime}}\times
$$
\begin{equation}
\label{slice}
\times \prod_{k=0} ^{N}
z_{k+1}^{*} \left( I + \frac{\Delta \beta }{\sqrt{E\beta _k}}\,
A(\beta_k) \right) z_k.
\end{equation}
Denoting $z_k = z(\beta_k)$ and taking into account that for
$\vert z_k \vert = 1$
$$
z^{*}_{k+1}z_k\,=\,1\,+\,(z^{*}_{k+1}-z^{*}_{k})z_k
$$
the product in (\ref{slice}) takes the form
$$
\prod_{k=0}^{N}\left(1\,+\,\dot{z}^{*}(\beta)z(\beta)\Delta \beta\,
+\,z^{*}(\beta +\Delta \beta)\frac{\Delta \beta }{\sqrt{E\beta _k}}A(\beta)
z(\beta)\right)\simeq
$$
$$
\simeq\,\exp\int_{x_B}^x(\dot{z}^{*}z+z^{*}
\frac{\Delta \beta }{\sqrt{E\beta}}A\,z)\,d\beta,
$$
where dot means the derivative over $\beta$. As a result the amplitude
is given by the functional integral over the complex vector field $z(\beta)$:
$$
T_{\mu \mu ^{\prime }}^{(n)}(x,y)\,=\,\sqrt{E}\,\frac xy \,\frac{\partial}
{\partial x}\,P_{\mu \mu ^{\prime }}^{(n)}(x,y),
$$
$$
P_{\mu \mu ^{\prime }}^{(n)}(x,y)=
c^{\prime}
\int Dz^{*}_nDz_n\delta (z^{*}_nz_n-1)\,
z_{n,\mu}(x)z^{*}_{n,\mu^{\prime}}(y)\times
$$
\begin{equation}\label{fint}
\times \exp
\left\{ \int_{y}^x d\beta \left[ \dot{z}^{*}_nz_n\, +\,
z^{*}_n \frac{A_n(\beta)}{\sqrt{E\beta}}z_n \right]\right\}.
\end{equation}
Here $c^{\prime}$ is the normalization constant and
\begin{equation} \label{measure}
Dz^{*}_nDz_n\delta (z^{*}_nz_n-1)\,=
\,\prod_{\beta}dz^{*}_n(\beta)dz_n(\beta)\delta (z^{*}_n(\beta)z_n(\beta)-1).
\end{equation}
Instead of an explicit $\delta$-function in measure (\ref{measure}) one can
rewrite the integral as
$$
P_{\mu \mu ^{\prime }}^{(n)}(x,y)=
c^{\prime\prime}\int D\sigma_n \int Dz^{*}_nDz_n\times
$$
\begin{equation}\label{fint2}
\times \exp
\left\{ \int_{y}^xd\beta \left[ \dot{z}^{*}_n z_n +
i\sigma_n (z^{*}_n z_n-1)
+z^{*}_n \frac{A_n(\beta)}{\sqrt{E\beta}}z_n)\right]\right\},
\end{equation}
where auxiliary field $\sigma_n(\beta)$ is introduced.
The functional representation
(\ref{fint}) or (\ref{fint2}) has to be substituted in correlator (\ref{corr}).
Averaging of exponential factors over vacuum state can again be done by making
use of formula (\ref{uv}) taking
$$
U_{n,i}(\beta)a_{n,i}(\beta)\,=\,\frac{1}{E\beta}\,
z^{*}_{n+1,\mu}(\beta)a_{n,\mu\nu}(\beta)z_{n+1,\nu}(\beta)
$$
$$
V_{n,i}(\beta)a_{n,i}^{\dagger}(\beta)\,=\,\frac{1}{E\beta}\,
z^{*}_{n,\mu}(\beta)a^{\dagger}_{n,\mu\nu}(\beta)z_{n,\nu}(\beta).
$$
It results into expression
\begin{equation}\label{Kint}
K_{\mu_1,\mu_1^{\prime},\ldots,\mu_N,\mu_N^{\prime}}(x_1,y_1,\ldots,x_N,y_N)=
\int\prod_{n}Dz^{*}_nDz_n \delta (z^{*}_nz_n-1)\,z_{n,\mu_n}(x_n)
\times
\end{equation}
$$
\times z^{*}_{n,\mu^{\prime}_n}(y_n)\, F(z^{*}_n,z_n)\,
\exp\left\{\sum_n \left[\int_{y_n}^{x_n} d\beta \dot{z}^{*}_n z_n\,
+\,\int_{y_{n,n+1}}^{x_{n,n+1}} d\beta \frac 1{E\beta}\, h_{n,n+1},
\right]\right\}
$$
$$
h_{ik}\,=\,z^{*}_{i,\mu} z^{*}_{k,\mu} z_{i,\nu} z_{k,\nu},
$$
where pre-exponential factor $F(z^{*},z)$ arises after differentiating the
exponent over $x_i$ variables:
\begin{equation}\label{Fzz}
F(z^{*},z)={\prod_i}^{\prime} \bigl[x_i z_i^{*}(x_i) z_i(x_i)+
h_{i-1,i}\theta(x_{i-1}-x_i)+h_{i,i+1}\theta(x_{i+1}-x_i) \bigr].
\end{equation}
Prime here denotes that after multiplication one has to replace
$$
\bigl( h_{ik} \theta(x_i-x_k)\bigr)^2 \rightarrow
h_{ik}\, x_i \delta(x_i-x_k).
$$

Within DLA accuracy all $y_i$ in the exponent have to be taken to be of the
order of $x_B$ while all $x_i \simeq x \sim 1$. After that
integral (\ref{Kint})can be treated as a functional
integral for a one-dimensional quantum mechanical chain.

Let $c^{\dagger}_{n,\lambda}$ and $c_{n,\lambda}$
be the creation and annihilation operators of the particles labelled by the
index $\lambda$ in the chain
site $n$ and let the dynamics of the system is described by the hamiltonian
$$
H\,=\,\sum_n c^{\dagger}_{n+1,\mu}c^{\dagger}_{n,\mu}\,c_{n+1,\nu}c_{n,\nu}
$$
which commutes with the operator of the number of particles in site $n$
$$
N_n\,=\,\sum_{\lambda=1}^2c^{\dagger}_{n,\lambda}c_{n,\lambda}.
$$

Writing $\delta$-functions
in the measure of Eq.(\ref{Kint}) through additional integrals over functions
$\sigma(\beta)$ the integral over $z^{*},z$ variables can be considered as a
functional integral in the holomorphic representation written for
the matrix element
\begin{equation}\label{matint}
M=
\end{equation}
$$
\left\langle 0_c\left|\prod_m c_{m,\mu_m}
F(c^{\dagger},c)\,\exp\left\{\int_{x_B}^x d\beta \bigl[\sum_n i\sigma_n(\beta)
(N_n-1)\,+\,\frac 1{E\beta} H \bigr]\right\}
\prod_n c_{n,\mu_n^{\prime}}^{\dagger} \right| 0_c\right\rangle,
$$
where $\left| 0_c\right\rangle$ is the vacuum state with respect
to operators $c_{n,\lambda}$. In the holomorphic representation operators
$c^{\dagger}_{n,\lambda}$ and $c_{n,\lambda}$ are replaced by $c$-number
functions $z^{*}_{n,\lambda}(\beta)$ and $z_{n,\lambda}(\beta)$ over which
the integration is performed (see e.g. \cite{FS}).
One can easily check it up by dividing the
''time'' interval $[x_B,x]$ into large number of small ones and repeating
step by step the derivation of the formula (\ref{fint}) inserting between
the parentheses the unit operator
\begin{equation}\label{I}
I\,=\,\int \prod_{n,\lambda} dz^{*}_{n,\lambda}dz_{n,\lambda}\,
e^{-\sum_{n,\lambda} z^{*}_{n,\lambda}z_{n,\lambda}}\left| Z\right\rangle
\left\langle Z \right|,
\end{equation}
where the state
$$
\left| Z\right\rangle\,=\,\exp\left\{\sum_{n,\lambda}z_{n,\lambda}
c^{\dagger}_{n,\lambda}\right\}\,\left| 0_c\right\rangle
$$
has the properties
$$
\begin{array}{c}
c_{n,\lambda}\left| Z\right\rangle\,=\,z_{n,\lambda}\left| Z\right\rangle\\
\left\langle Z_2 \left|\right. Z_1 \right\rangle\,=\,e^{-z^{*}_2 z_1}\\
\left\langle Z_2 \left|H(c^{\dagger},c)\right| Z_1 \right\rangle\,=\,
H(z^{*}_2,z_1)e^{-z^{*}_2 z_1}
\end{array}
$$
and operator $H$ is supposed to be normal ordered.

The functional integral over $\sigma(\beta)$ gives $N_n=1$ in (\ref{matint}),
i.e. the nonvanishing contribution to the matrix
element comes only from the states where there is only one particle in each
chain site $n$. Acting on these states hamiltonian $H$ transforms them into
themselves since it does not change the total number of particles in the site.
Finally, correlator (\ref{Kint}) is expressed through the matrix element in
the kinematic where $x_i \simeq x,\, y_i \simeq x_B$ by the formula
$$
K_{\mu_1,\mu_1^{\prime},\ldots,\mu_N,\mu_N^{\prime}}(x_1,\ldots,x_N)=
$$
\begin{equation}\label{matfin}
=<0_c| \prod_m c_{m,\mu_m}F(c^{\dagger},c)\,e^{\frac 1E \ln \frac x{x_B} H}
\,\prod_n c_{n,\mu_n^{\prime}}|0_c>.
\end{equation}
The powers of $z_i^{*}(x_i) z_i(x_i)$ in pre-exponential factor
$F(z^{*},z)$ (\ref{Fzz}) do not contribute to functional
integral (\ref{Kint}) (one can treat them as a result of insertion expression
(\ref{I}) between unit operators) therefore
\begin{equation}\label{Fcc}
F(c^{*},c)\,=\,N\left\{{\prod_i}^{\prime} \bigl[
H_{i-1,i}\theta(x_{i-1}-x_i)+H_{i,i+1}\theta(x_{i+1}-x_i) \bigr]\right\},
\end{equation}
where
$$
H_{ik}\,=\,c^{\dagger}_{i,\mu} c^{\dagger}_{k,\mu} c_{i,\nu} c_{k,\nu},
$$
\begin{equation}\label{H}
H\,=\,\sum_n H_{n,n+1},
\end{equation}
prime has the same meaning as in Eq.(\ref{Fzz}) and $N$ means the normal
ordering.

Introducing helicity basis $e^{\pm }=\frac 1{\sqrt{2}}
\left( 1,\pm i\right)$ the two-body operators $H_{ik}$ act on the chain
wavefunction $\psi$ as
\begin{equation}
\label{hel}
\begin{array}{c}
(H_{ik}\psi )_{+-}=(\psi _{+-}+\psi _{-+}) \\
(H_{ik}\psi )_{-+}=(\psi _{-+}+\psi _{+-}) \\
(H_{ik}\psi )_{++}=(H_{ik}\psi )_{--}=0,
\end{array}
\end{equation}
where all indexes except those at $i,k$ positions are omitted. Introducing
for each chain site an effective spin-$\frac12$ variable $s_z$ and treating
the positive helicity as $s_z=\frac12$ state and the negative one
as $s_z=-\frac12$ the term $H_{ik}$ takes a form
$$
H_{ik}\,=\,\left(S^i_{+}S^k_{-}\,+\,S^i_{-}S^k_{+}\right)\,-\,
\left(S^i_{z}S^k_{z}\,-\,\frac 14 \right),
$$
where $S_{+,-,z}$ are the spin operators. Thus the total chain hamiltonian
(\ref{H}) turns out to be the hamiltonian of Ising model for one-dimensional
magnets (for closed chain).

However there is an uncertainity in this way of proceeding, namely, a general
normalization constant that usually remains to be unknown in the functional
integral approach. As a consequence hamiltonian $H$ in (\ref{corr}) is
determined up to constant $E_0$ which leads to the shift of the spectrum.
To find $E_0$ one has to compare the result with a case where the answer is
known since $E_0$ depends only on the way the functional integral is built up
rather than a particular structure of the interaction. To this purpose the
simple kernel
$$
\label{kern}\Phi_{\mu ,\nu ,\lambda ,\sigma }(\beta _1,\beta _2\mid
\beta _1^{\prime },\beta _2^{\prime })\,=\,\delta _{\mu ,\lambda
}\delta _{\nu ,\sigma }\,\beta _1\delta (\beta _1-\beta _2)\,\frac
1{\beta _1^{\prime }\beta _2^{\prime }}.
$$
can be taken instead of (\ref{kern1}). This kernel is diagonal in polarization
indexes, therefore the solution is given by expression (\ref{simp}). From the
other hand it can be obtained through all functional integral machinery
described above that leads to the quantum chain with hamiltonian
$$
H\,=\,\sum_n c^{\dagger}_{n+1,\mu}c_{n+1,\mu}\,c^{\dagger}_{n,\nu}c_{n,\nu}.
$$
Due to the restriction $N_n=1$
$$
H\,=\,\sum_n 1\,=\,N
$$
and the asymptotic behaviour of the matrix element is the same as correlator
(\ref{simp}) has. Thus one can conclude that energy shift $E_0=0$.

{\bf 5.} The solution of the Ising model is well known.
After substituting the wavefunction of the Schr\"odinger equation
$$
H\,\psi^{(N)}_m\,=\,\varepsilon^{(N)}_m\,\psi^{(N)}_m\
$$
the asymptotic behaviour of matrix element (\ref{matfin}) is defined by the
maximal energy $\varepsilon^{(N)}_{max}$. Then
$$
W(E)\,\sim\,\exp\left\{\frac1E\,\varepsilon^{(N)}_{max}\,\ln\frac{x}{x_B}
\right\}
$$
and
$$
\rho (\xi )\,\sim\,\exp\,\sqrt{4\varepsilon^{(N)}_{max}\,\xi\ln\frac{x}{x_B}}.
$$

In the Ising model the zero-energy state (vacuum) is
the state where all spins have
the same orientation (e.g. when all they downward directed).
An excited state with $m$ turned-off spins is treated as the state with $m$
quasiparticles $(m<N)$ which are characterized by momenta $k_i$. The excitation
energy is given then by the formula
\begin{equation}\label{Ie}
\varepsilon^{(N)}_{m}\,=\,\sum_{i=1}^{m}\bigl(\cos k_i\,+\,1\bigr),
\end{equation}
where $k_i$ (or, more exactly, $e^{ik_i}$) values
are determined (for the closed chain case) by the
algebraic equation of high order (see e.g. \cite{gaudin}) so there is not
an analityc expression for them. Nevertheless Eq.(\ref{Ie}) shows that
$\varepsilon^{(N)}_{max}<2(N-1)$ (the state with $m=N$ obviously has zero
energy). For large $N$ one can expect the maximum energy for
$m \simeq \frac N2$ which leads to
$\varepsilon^{(N)}_{max}=s N$, where $s \sim 1$. Neglecting $\alpha_S$ running
it gives for the structure function
$$
x_BG_{2N}(x_B,Q^2)\sim \left( \frac {\mu ^2}{Q^2}\right)^N\exp
\sqrt{2sN\,\frac{\alpha_S N_c}{\pi}\,\ln Q^2/\mu ^2\ln 1/x_B}.
$$
Thus the contribution of a single $2N$-particle cylinder is smaller than
$N$ Pomerons one (when $N_c \rightarrow \infty$ Pomeron can be treated as
a two-particle cylinder). Qualitative agreement of this asymptotics with
formula (\ref{simp}) shows the incorporation of gluon polarization
does not change the behaviour of the structure function.
This is a consequense of small number gluons
interactions allowed by cylinder topology. The stronger $N$-dependence
is to be expected for $\frac 1{N_c}$ corrections. Therefore the investigation
of the high twist evolution equation for finite $N_c$ is of great interest.

\vspace{1cm}

{\bf Acknowledgements:} The author is grateful to J.Bartels and M.Ryskin
for helpful discussions.
He is also gratefully acknowledges the hospitality of DESY
and the financial support of the Volkswagenstiftung.

\end{document}